\documentclass[twocolumn,groupedaddress,prl,floatfix,superscriptaddress]{revtex4-1}

\usepackage{graphicx}
\usepackage{array}
\usepackage{amsmath}
\usepackage{bm}
\usepackage[colorlinks,allcolors=blue]{hyperref}

\usepackage{color}
\definecolor{CF}{rgb}{0.5, 0.1, 0.4}
\usepackage[normalem]{ulem}

\begin{document}

\title{\textit{Ab initio} vibrational free energies including anharmonicity for multicomponent alloys}
\author{Blazej Grabowski}
\affiliation{Department for Computational Materials Design, Max-Planck-Institut f\"ur Eisenforschung GmbH, Max-Planck-Str. 1, 40237 D\"usseldorf, Germany}
\author{Yuji Ikeda}
\affiliation{Department for Computational Materials Design, Max-Planck-Institut f\"ur Eisenforschung GmbH, Max-Planck-Str. 1, 40237 D\"usseldorf, Germany}
\author{Fritz K\"ormann}
\affiliation{Department for Computational Materials Design, Max-Planck-Institut f\"ur Eisenforschung GmbH, Max-Planck-Str. 1, 40237 D\"usseldorf, Germany}
\affiliation{Department of Materials Science and Engineering, Delft University of Technology, Mekelweg 2, 2628 CD Delft, Netherlands}
\author{Christoph Freysoldt}
\affiliation{Department for Computational Materials Design, Max-Planck-Institut f\"ur Eisenforschung GmbH, Max-Planck-Str. 1, 40237 D\"usseldorf, Germany}
\author{Andrew Ian Duff}
\affiliation{Scientific Computing Department, STFC Daresbury Laboratory, Hartree Centre, Warrington, UK}
\author{Alexander Shapeev}
\affiliation{Skolkovo Institute of Science and Technology, Skolkovo Innovation Center, Nobel St. 3, Moscow 143026, Russia}
\author{J$\rm{\ddot{o}}$rg Neugebauer}
\affiliation{Department for Computational Materials Design, Max-Planck-Institut f\"ur Eisenforschung GmbH, Max-Planck-Str. 1, 40237 D\"usseldorf, Germany}
\date{\today}

\begin{abstract}
A density-functional-theory based approach to efficiently compute numerically exact vibrational free energies---including anharmonicity---for chemically complex multicomponent alloys is developed. It is based on a combination of thermodynamic integration and a machine-learning potential. We demonstrate the performance of the approach by computing the anharmonic free energy of the prototypical five-component VNbMoTaW refractory high entropy alloy.
\end{abstract}

\maketitle
	
Recent developments in the field of multicomponent alloys [high entropy alloys (HEAs), compositionally complex alloys (CCAs)] have opened new materials design perspectives \cite{gludovatz2014fracture,li2016metastable,MYW14,GYW16}. The prediction and exploration of thermodynamic properties and phase stabilities is therefore of critical importance. To this end, parameter-free \textit{ab initio} calculations, particularly using density-functional theory (DFT), are rapidly gaining popularity \cite{IGK2018}. However, the requirement \cite{engin2008characterization,GSH2011,ZGN2017} to accurately capture small free energy differences ($\approx 1$\,meV/atom) poses severe challenges. Only very recently the required tools to accurately compute free energies of selected unary and ordered binary systems have been developed \cite{Duf2015,GGH2015,ZGN2017}, while efforts to treat the immense chemical complexity of multicomponent alloys are still in their infancy. Here, we propose a highly efficient approach to compute free energies of such multicomponent alloys. We apply it to a prototypical five-component equiatomic body-centered cubic (bcc) refractory VNbMoTaW HEA in its solid solution. This alloy has attracted attention for its superior high-temperature mechanical properties \cite{SENKOV2011698,SENKOV20101758}.

The free energy is determined by various mechanisms including electronic excitations, configurational entropy and atomic vibrations (e.g., Refs. \cite{ZGH2018,IGK2018}). The major contribution is due to atomic vibrations (e.g., for VNbMoTaW \cite{KIG17}: $8.6\,k_{\rm B}$ at 1500\,K vs. a configurational entropy of $1.6\,k_{\rm B}$), the leading term of which can be captured by the quasiharmonic approximation. However, the latter accounts only for the phonon softening due to volume expansion and misses out the temperature-dependent phonon softening and broadening. Effective harmonic Hamiltonians \cite{Sou2008,Sou2009,HSA2013,PhysRevB.89.064302,PhysRevB.89.094109,CARRERAS2017221} can approximately account for the temperature induced changes. Numerically exact vibrational free energies can be obtained by thermodynamic integration \cite{Alf2001,Alf2002-1,Alf2002-2,Lid2003,Mou2017},
\begin{equation}
  F = F^{\rm ref} + \int_0^1 d\lambda \langle E^{\rm DFT}-E^{\rm ref} \rangle_{\lambda}\label{TIeq},
\end{equation}
from a reference potential $E^{\rm ref}$ with free energy $F^{\rm ref}$ to DFT energies $E^{\rm DFT}$, where $\langle \cdots \rangle_{\lambda}$ denotes a thermal average on a mixed potential \mbox{$E^{\lambda}=\lambda E^{\rm DFT} + (1-\lambda)E^{\rm ref}$}. Using a harmonic reference would in principle give the exact anharmonic free energy including temperature-dependent phonon softening and broadening, but such a brute-force integration is computationally prohibitive in practice. The computational effort is dominated by: (1) the number of molecular-dynamics (MD) steps needed to obtain a statistically converged average, (2) the number of $\lambda$-values required to calculate the integral, and (3) the computational effort per MD step.

A state-of-the-art method, making the three steps more feasible, is the \textit{two-stage upsampled thermodynamic integration using Langevin dynamics} (TU-TILD) method \cite{Duf2015}, which employs an interatomic potential as an intermediate reference in the thermodynamic integration. The thermodynamic integration is split into two stages, first from the harmonic to the reference potential, secondly from the reference potential to full DFT. The intention is to reduce the number of steps necessary to converge the thermal average in the second stage---containing the explicit, costly DFT calculations---by fitting the potential as closely as possible to the DFT data. The brunt of the statistical convergence is then relegated to the thermal average in the first stage, which does not contain explicit DFT calculations and can be thus computed highly efficiently.

\begin{figure*}[ht]\centering
 \resizebox{.9\textwidth}{!}{ \includegraphics{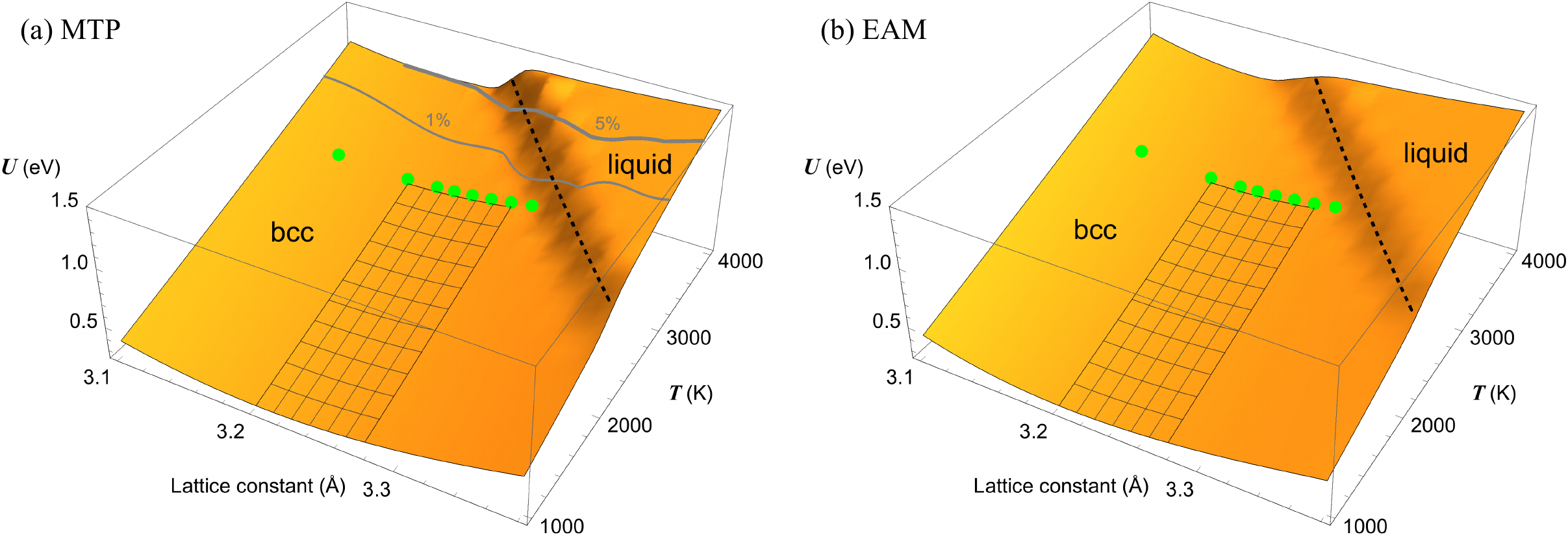} }
 \caption{Internal energy, $U$, maps as a function of lattice constant/volume and temperature, $T$, for (a) MTP and (b) EAM. Green dots mark the volumes (at 3000 K) used for fitting the potentials. Black dotted lines emphasize the transition to the liquid phase. The squared volume and temperature region is the region of interest (3.20\,...\,3.28 \AA{} in terms of the lattice constant). The gray lines in (a) are contours at which 1 or 5\% of the MTP MD runs are unstable \cite{supp}, however, all the runs can be made stable by using \emph{active learning} \cite{podryabinkin2016-active-learning}.\label{U}}
\end{figure*}

The performance of this approach relies critically on the feasibility of fitting a potential that accurately interpolates the DFT data within the thermally accessible phase space. For HEAs and CCAs, where the primary goal is the exploration of the large compositional space and thus many different atomic structures, the requirement of an efficient reference for thermodynamic integration is even more critical. At the same time, for multicomponent alloys the number of fitting parameters drastically increases which results in a serious challenge to fitting reliable potentials. {\it A priori} it is not clear whether such an approach is at all feasible for HEAs and CCAs. 

A possible solution to this problem could be offered by the emerging class of machine-learning techniques which have recently been developed in various scientific fields \cite{rupp2018-ml-chemistry-editorial}. Several machine-learning potentials have been proposed so far \cite{PhysRevLett.98.146401,doi:10.1063/1.3553717,PhysRevLett.104.136403,Bartoke1701816,PhysRevB.92.054113,PhysRevMaterials.1.063801}. For example, Gaussian process regression was applied to approximate the potential free energy surface of small and medium-sized molecules across the slow degrees of freedom \cite{stecher2014-free-energy-peptides}. First attempts have been put forward to describe alloys \cite{2018arXiv180610567G}, focussing on the configurational degree-of-freedom of a ternary system. Whether a machine-learning approach is applicable to compute the vibrational free energy of chemically even more complex bulk materials is so far unknown.

In this work we develop a new algorithm combining the TU-TILD method with moment tensor potentials (MTP), a class of machine-learning potentials first proposed in Ref.~\cite{shapeev2016mtp}, and recently shown to perform best among different machine-learning models \cite{comp}. We demonstrate here that the TU-TILD+MTP combination is an ideal symbiosis for an efficient and accurate calculation of the full vibrational free energy of disordered multicomponent alloys. MTP describes the atomic environment of the $i$th atom by the moments of inertia of the neighboring atoms,
\[
M_{n,m} = \sum_{j} f_{n,i,j}(r_{ij}) \underbrace{{\bm r}_{ij}\otimes {\bm r}_{ij}\otimes\ldots\otimes {\bm r}_{ij}}_{m\text{ times}}.
\]
Here the radial functions $f_{n,i,j}(r_{ij})$, $n=1,2,\ldots$, define different shells around the $i$th atom; the contribution of the $j$th atom to the $n$th shell can depend on the types of the $i$th and $j$th atoms. When $m=0$, $M_{n,0}$ is a scalar quantity interpreted as the weight of the $n$th shell. The set of these scalar descriptors is not complete. However, this set can be made complete by adding vectorial ``eccentricity'' of the $n$th shell, the tensor of second moments of inertia $M_{n,2}$, the third moments $M_{n,3}$, etc. Hence, MTP can approximate an arbitrary local interaction energy by forming basis functions as different ways of contracting these tensor-valued moment descriptors to a scalar and considering an arbitrary linear combination of these basis functions with parameters fitted from data \cite{shapeev2016mtp,2018arXiv180610567G}. In practice, this means that we can increase the number of parameters until the fitting error stops decreasing---this would indicate that we have reached a lower error bound that a local model can achieve with a given cutoff radius.

We now apply an MTP as a reference potential within TU-TILD for the disordered VNbMoTaW HEA. Chemical disorder is modeled by a special quasirandom structure (SQS) in a 125 atomic supercell \cite{Zunger_PRL_1990_Special}. DFT calculations are performed with \textsc{vasp} \cite{KF1996,KF1996b} employing PAW \cite{Blo1994}, GGA-PBE \cite{PBE1996}, and including electron-phonon coupling \cite{Zhang2017}. See the supplement for details \cite{supp}. We consider temperatures up to 3000 K which is close to the estimated melting point \cite{SENKOV20101758}. In accord with our previous works \cite{ZGN2017,GGG2018} we use DFT MD simulations at several volumes at a high temperature to provide sufficient fitting data for MTP [green dots in Fig.~\ref{U}(a)]. We choose a cutoff radius of $5$ \AA\ (including the first up to the third neighbor shell). Additional tests for a smaller cutoff including two shells show a small change \cite{supp}.

Since machine-learning potentials have an inherently low extrapolation capacity, stability over the relevant part of the phase space is a critical issue. Detailed tests reveal that the MTP, fitted according to the above procedure, is sufficiently stable in the relevant volume and temperature range for the application within TU-TILD [see Fig.~\ref{U}(a)]. In fact the potential can be also used at extrapolated volumes and temperatures and predicts even the onset of the liquid phase. Only a small number of MD runs in the range of a few \% (see gray contour lines) becomes unstable. The results shown in Fig.~\ref{U}(a) correspond to a ``single-shot'' MTP potential fitted to an initial set of DFT data. However, the MTP provides an inherent metric \cite{podryabinkin2016-active-learning} to quantify the degree of extrapolation and thus offers the possibility to actively sample configurations for fitting, ensuring stability for the temperatures of interest.

The performance of the MTP as a reference potential within TU-TILD can be quantified by: (1) The dependence of $\langle E^{\rm DFT}-E^{\rm ref}\rangle_\lambda$ on $\lambda$ [Eq.~(\ref{TIeq}); where $E^{\rm ref}$ stands for MTP energies] should be as smooth as possible, (2) the standard deviation of the energy difference $E^{\rm DFT}-E^{\rm ref}$ should be as small as possible, and (3) the correlation in the forces should be as strong as possible. The thermodynamic integration from the harmonic potential to MTP will not be discussed since, as mentioned above, this stage of the integration can be computed highly efficiently given the fact that the MTP is more than six orders of magnitude faster than DFT. See the supplement for detailed timings \cite{supp}.

\begin{figure}[t!]\centering
 \resizebox{.9\columnwidth}{!}{ \includegraphics{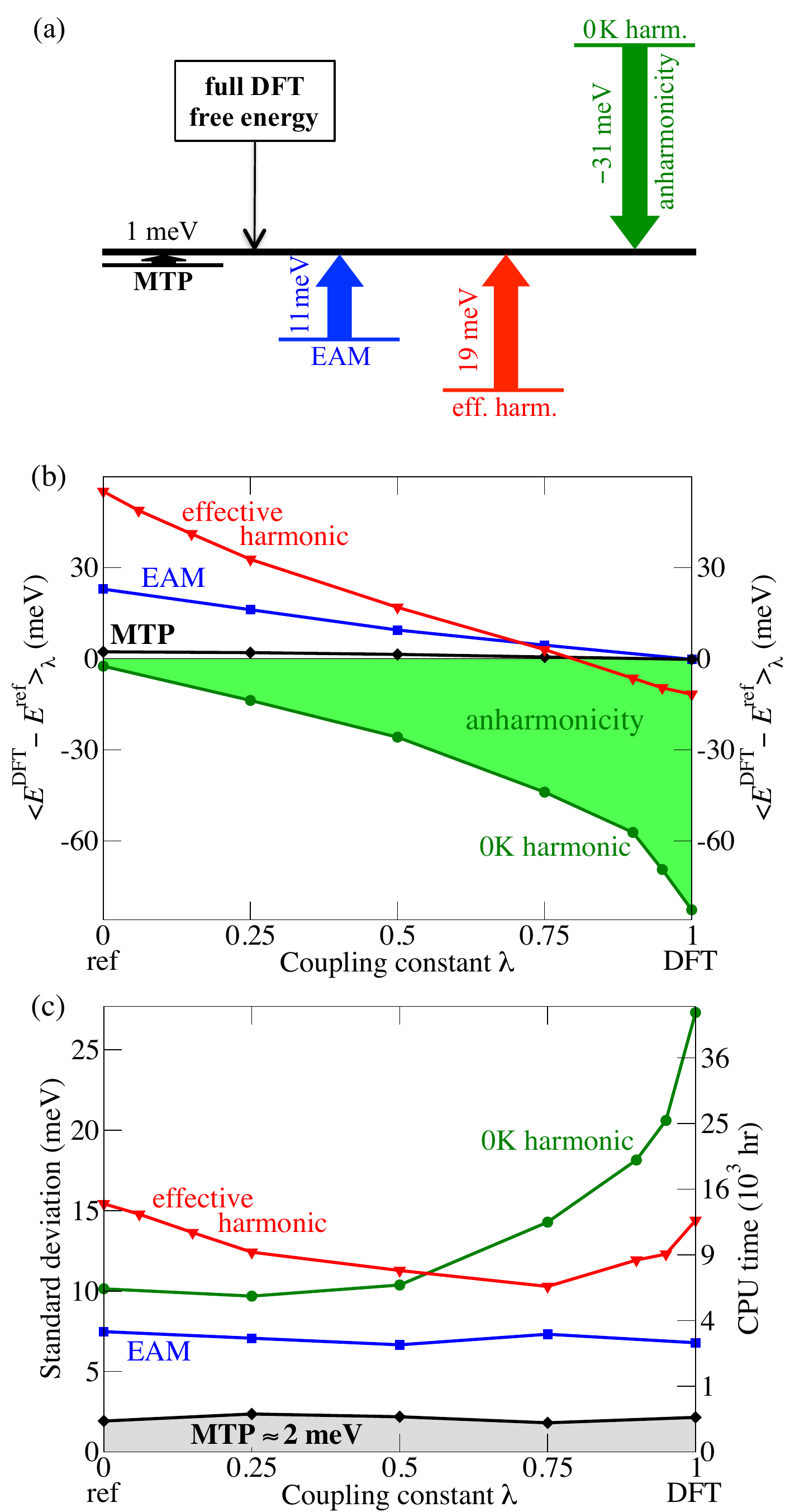} }
 \caption{Results of thermodynamic integration to DFT for VNbMoTaW using different references at 3000 K. The integral over the curves in (b) gives the difference in free energy between DFT and the reference shown in (a). The smaller the standard deviation shown in (c), the quicker the statistical convergence of the curves in (b) as indicated on the right axis in (c). For the CPU time calculation a standard error of 1 meV/atom, a CPU time of 4 hr per ionic step, and a correlation length of 15 steps were taken.\label{TI}}
\end{figure}

\begin{figure*}[ht]\centering
 \resizebox{1\textwidth}{!}{ \includegraphics{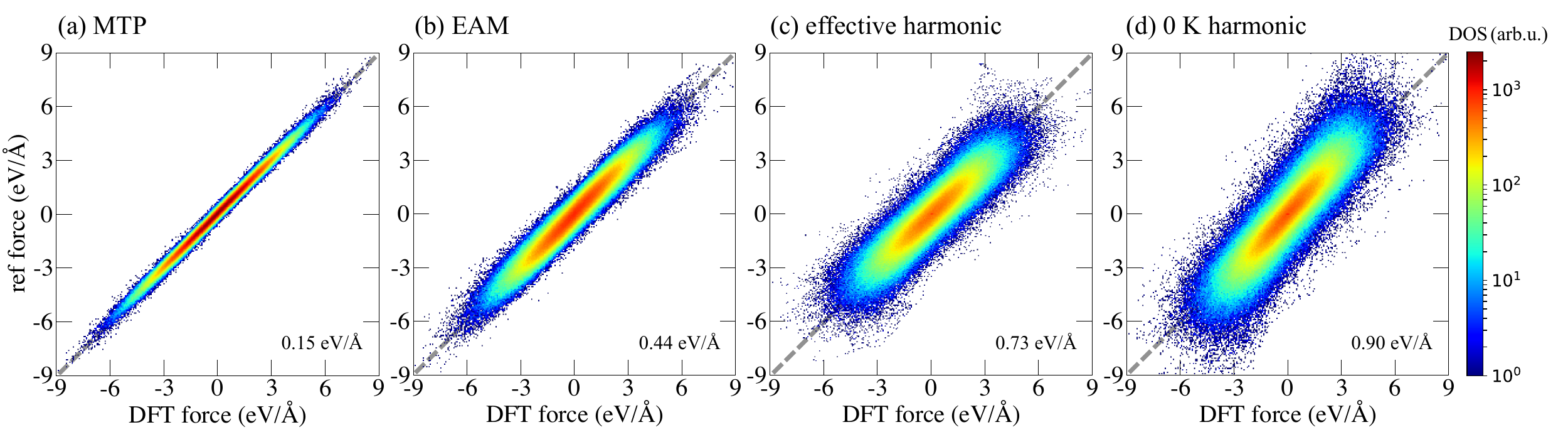} }
 \caption{(Color online) Correlation of the DFT forces \textit{versus} forces from the different approximations at 3000 K. The color represents to the local density. The numbers in the right lower corners represent the root mean square error of the distributions. \label{f}}
\end{figure*}

The excellent performance of the MTP is demonstrated in Figs.~\ref{TI} and \ref{f}(a). The MTP energies are so close to the DFT energies that the thermal average $\langle E^{\rm DFT}-E^{\rm ref}\rangle_\lambda$ is almost independent of $\lambda$ and close to the targeted error of 1 meV [Fig.~\ref{TI}(b), black curve], i.e., the resulting MTP \textit{free} energy is only 1 meV/atom away from the DFT free energy [Fig.~\ref{TI}(a)]. Computing this difference can be done highly efficiently because of the small standard deviation in the range of only 2 meV/atom [Fig.~\ref{TI}(c)]. Consistently, the MD forces predicted by the optimized MTP show a strong correlation with the DFT forces [Fig.~\ref{f}(a)]. This good performance of the MTP is found for the whole relevant volume range \cite{supp} demonstrating that an efficient study of the thermal expansion is possible as well. 

To set a baseline for the performance of the MTP we employ alternative reference potentials that have been used previously for chemically less complex unary and binary systems: (1) 0\,K harmonic \cite{Gra2009,GGH2015,GWG2015}, (2) effective harmonic \cite{HSA2013}, and (3) embedded atom method (EAM) \cite{Duf2015,GGG2018}. We start with the 0\,K harmonic potential computed for VNbMoTaW in Ref.~\cite{KIG17}. As mentioned above, using this reference in Eq.~(\ref{TIeq}) directly provides the anharmonic contribution.

Figure \ref{TI} highlights the difficulties. Figure \ref{TI}(b) shows the very nonlinear dependence of the thermodynamic average, $\langle E^{\rm DFT}-E^{\rm ref}\rangle_\lambda$ (where $E^{\rm ref}$ stands now for the harmonic energies), on the coupling constant $\lambda$ (green curve). Due to this nonlinear behavior the evaluation requires many sampling points. Using the MTP reference introduced above makes a highly accurate calculation of this quantity possible. We find that at 3000 K the anharmonic contribution is $-31$ meV/atom. Due to their more open structure, bcc materials are known to have a more complex temperature dependence of the anharmonic free energy than close-packed materials \cite{Ful2010}. Our calculations confirm this observation quantitatively: The anharmonic contribution is almost twice as large as for previously investigated, close-packed face-centered cubic (fcc) structures (range of 1 to 25~meV/atom at the melting point) \cite{GGH2015}.

An even more serious issue than the strongly nonlinear dependence of \mbox{$\langle E^{\rm DFT}-E^{\rm ref}\rangle_\lambda$} is the fact that long MD runs are required to statistically converge each of these points. The underlying reason is that the standard deviation of the energy difference, $E^{\rm DFT}-E^{\rm ref}$, is large as shown in Fig.~\ref{TI}(c). The large difference in the energies is also reflected in the weak correlation between the harmonic and DFT forces during an MD run as shown in Fig.~\ref{f}(d).

We now investigate an effective harmonic force constant matrix constructed at finite temperature as a reference. Such a matrix is employed, e.g., in the temperature-dependent effective potential (TDEP) method \cite{HSA2013}, and provides the advantage that analytical formulas can be used to compute the vibrational free energy. Our tests show that including pair interactions up to the first- and second-nearest neighbors gives similar results as with an additional third shell \cite{supp}. Results for three shells will be discussed in the following. The interactions are determined from a least-square fit of the forces from more than 1500 configurations of an MD run at 3000\,K at the target lattice constant. Owing to the harmonic approximation, the fitting problem is linear \cite{Freysoldt2017} and is solved with a standard algebraic method, namely the pseudo-inverse from singular value decomposition to avoid accidental ill-conditioning. The zero-force reference structure, where the potential attains its minimum is set to the $0$\,K equilibrium geometry.

The forces from such an effective harmonic potential show a slightly better correlation with the DFT MD forces at the target temperature than the 0\,K harmonic forces [cf. Figs. \ref{f}(c) vs. (d)]. Correspondingly the dependence of $\langle E^{\rm DFT}-E^{\rm ref}\rangle_\lambda$ on $\lambda$, where $E^{\rm ref}$ now stands for the effective harmonic energies, is less nonlinear than for the 0\,K harmonic energies [Fig. \ref{TI}(b), red vs. green]. However, the standard deviation is smaller only at $\lambda>0.5$. Thus an effective harmonic potential offers only a slightly improved reference for thermodynamic integration ($\approx$ 1.5 times faster). The respective vibrational free energy obtained using the standard harmonic formulas including a correction of the internal energy as introduced in the TDEP method \cite{HSA2013} gives a slightly reduced error of 19 meV/atom compared to $-31$ meV/atom for the 0\,K harmonic reference [Fig.~\ref{TI}(a)].

The still rather large error of the effective harmonic matrix is related to strong local pairwise anharmonicity \cite{GGH2015}. The inherent asymmetry of the nearest neighbor potential when atoms move together or apart cannot be captured in general by any harmonic potential irrespective of the temperatures it is fitted to. To take the required asymmetry properly into account an asymmetric potential parametrization is required. Asymmetric potentials are offered by the MTP discussed already above or likewise by an EAM parametrization.

We thus investigate whether an EAM fit for the complex disordered VNbMoTaW HEA is possible and, if it is, how it performs for thermodynamic integration. We employ the \textsc{meamfit v2} package \cite{DUFF2015439,meamfit}. Our tests show that the number of expansion parameters has only a small influence \cite{supp}. Results shown in the following refer to our best EAM with 3 embedding terms per species, with 11 parameters for the electron densities, and 19 for the pair-potentials. In total there are 355 independent parameters for this chemically highly complex quinary system, rendering the extraction of an accurate potential particularly challenging. To address this challenge we fit initially to a subset of the $\approx 8000$ energies available across all volumes. This subset consists of $\approx 2000$ uniformly-spaced energies, providing sufficient points per parameter to prevent over-fitting. We then take the best performing potential as a starting point for a single shot conjugate gradient fit to {\sl all} $\approx 8000$ energies.

During the optimization, energies are computed relative to the 0\,K relaxed structure for the corresponding volume. A cutoff radius of 5\,{\AA} (as for MTP) is imposed for the pairwise terms, and negative `electron' densities are allowed---although positive background densities are required overall---to provide maximum variational flexibility. The resulting potential is stable across a wide range of volumes and temperatures [see Fig.~\ref{U}(b)] and predicts the onset of the liquid phase.

The forces obtained with the fitted EAM potential show a better correlation with DFT MD forces [Fig.~\ref{f}(b)] than the harmonic potentials. Consistently, the dependence of $\langle E^{\rm DFT}-E^{\rm ref}\rangle_\lambda$ on $\lambda$, where $E^{\rm ref}$ stands now for the EAM energies, is more linear and the standard deviation is smaller for all $\lambda$ (blue curves in Fig.~\ref{TI}). Using the EAM as a reference is more than three times faster than an effective harmonic potential. However, the EAM cannot compete with the MTP as a reference potential which further increases the efficiency by about an order of magnitude as compared to the EAM.

These results reveal that the combination of TU-TILD with MTP represents the presently most efficient combination to compute the vibrational free energy contribution. A preliminary study \cite{prelim} shows that this does not only apply to equiatomic compositions such as the one studied in the present manuscript but likewise to arbitrary non-equiatomic compositions, and further also to different crystallographic lattice types such as hcp or fcc, magnetic systems, and even to the liquid phase.

The underlying physical reason for the excellent performance of the TU-TILD+MTP combination is the fact that the vibrational free energy is determined by a rather well-defined, sufficiently smooth, and local---although strictly anharmonic---part of the phase space. Several other studies \cite{GGH2015,Zhang2017,SFE} have already indicated that long-ranged interactions that maybe present at $T=0$\,K due to quantum-mechanical interference effects, vanish when explicit vibrations are introduced at finite temperatures due to the breaking of the crystal symmetries. Effective interactions at finite temperatures are thus localized and can be well fitted by a local approach. These interactions are strongly anharmonic requiring an anharmonic description as provided by the MTP or EAM, with a greater flexibility offered by the MTP \cite{foot}. It should be stressed that this greater flexibility comes with a lower extrapolation capability (see Fig.~\ref{U} and corresponding discussion). A main achievement of the present work is having shown that, for free energy calculations within the TU-TILD+MTP approach, the stability of the MTP is sufficient and that providing a set of well distributed fitting data renders the sampling of the thermally accessible phase space an interpolation task; a task optimally suited for a machine learning approach.

To open the approach to a broad community we are presently implementing it into the pyiron environment \cite {pyiron-paper,py}. This, together with the performance of the new approach, paves the route to compute vibrational free energies not only highly accurately but with a computational performance adequate for high-throughput screening of multicomponent alloys.

{\bf Acknowledgments}
We thank Jan Janssen and Konstantin Gubaev for fruitful discussions. Funding by the Deutsche Forschungsgemeinschaft (SPP 2006) and the European Research Council (ERC) under the EU's Horizon 2020 Research and Innovation Programme (Grant No. 639211) is gratefully acknowledged. F.K. was supported by the NWO/STW (VIDI grant 15707). A.S. was supported by the Russian Science Foundation (Grant No 18-13-00479). A.I.D. acknowledges support from the STFC Hartree Center program, Innovation: Return on Research, funded by the UK Department for Business, Energy \& Industrial Strategy.

{\bf Competing interests}
The Authors declare no Competing Financial or Non-Financial Interests.

{\bf Data availability}
The authors declare that all data supporting the findings of this study are available within the paper and its supplementary information files.

{\bf Contributions}
B.G. performed the TU-TILD approach, C.F. the implementation of the effective harmonic force constant fitting procedure, A.I.D. the MEAM-fits, A.S. developed and fitted the MTPs. All authors designed the project, discussed the results, and wrote the manuscript.


\begin{thebibliography}{51}%
\makeatletter
\providecommand \@ifxundefined [1]{%
 \@ifx{#1\undefined}
}%
\providecommand \@ifnum [1]{%
 \ifnum #1\expandafter \@firstoftwo
 \else \expandafter \@secondoftwo
 \fi
}%
\providecommand \@ifx [1]{%
 \ifx #1\expandafter \@firstoftwo
 \else \expandafter \@secondoftwo
 \fi
}%
\providecommand \natexlab [1]{#1}%
\providecommand \enquote  [1]{``#1''}%
\providecommand \bibnamefont  [1]{#1}%
\providecommand \bibfnamefont [1]{#1}%
\providecommand \citenamefont [1]{#1}%
\providecommand \href@noop [0]{\@secondoftwo}%
\providecommand \href [0]{\begingroup \@sanitize@url \@href}%
\providecommand \@href[1]{\@@startlink{#1}\@@href}%
\providecommand \@@href[1]{\endgroup#1\@@endlink}%
\providecommand \@sanitize@url [0]{\catcode `\\12\catcode `\$12\catcode
  `\&12\catcode `\#12\catcode `\^12\catcode `\_12\catcode `\%12\relax}%
\providecommand \@@startlink[1]{}%
\providecommand \@@endlink[0]{}%
\providecommand \url  [0]{\begingroup\@sanitize@url \@url }%
\providecommand \@url [1]{\endgroup\@href {#1}{\urlprefix }}%
\providecommand \urlprefix  [0]{URL }%
\providecommand \Eprint [0]{\href }%
\providecommand \doibase [0]{http://dx.doi.org/}%
\providecommand \selectlanguage [0]{\@gobble}%
\providecommand \bibinfo  [0]{\@secondoftwo}%
\providecommand \bibfield  [0]{\@secondoftwo}%
\providecommand \translation [1]{[#1]}%
\providecommand \BibitemOpen [0]{}%
\providecommand \bibitemStop [0]{}%
\providecommand \bibitemNoStop [0]{.\EOS\space}%
\providecommand \EOS [0]{\spacefactor3000\relax}%
\providecommand \BibitemShut  [1]{\csname bibitem#1\endcsname}%
\let\auto@bib@innerbib\@empty
\bibitem [{\citenamefont {Gludovatz}\ \emph {et~al.}(2014)\citenamefont
  {Gludovatz}, \citenamefont {Hohenwarter}, \citenamefont {Catoor},
  \citenamefont {Chang}, \citenamefont {George},\ and\ \citenamefont
  {Ritchie}}]{gludovatz2014fracture}%
  \BibitemOpen
  \bibfield  {author} {\bibinfo {author} {\bibfnamefont {B.}~\bibnamefont
  {Gludovatz}}, \bibinfo {author} {\bibfnamefont {A.}~\bibnamefont
  {Hohenwarter}}, \bibinfo {author} {\bibfnamefont {D.}~\bibnamefont {Catoor}},
  \bibinfo {author} {\bibfnamefont {E.~H.}\ \bibnamefont {Chang}}, \bibinfo
  {author} {\bibfnamefont {E.~P.}\ \bibnamefont {George}}, \ and\ \bibinfo
  {author} {\bibfnamefont {R.~O.}\ \bibnamefont {Ritchie}},\ }\href@noop {}
  {\bibfield  {journal} {\bibinfo  {journal} {Science}\ }\textbf {\bibinfo
  {volume} {345}},\ \bibinfo {pages} {1153} (\bibinfo {year}
  {2014})}\BibitemShut {NoStop}%
\bibitem [{\citenamefont {Li}\ \emph {et~al.}(2016)\citenamefont {Li},
  \citenamefont {Pradeep}, \citenamefont {Deng}, \citenamefont {Raabe},\ and\
  \citenamefont {Tasan}}]{li2016metastable}%
  \BibitemOpen
  \bibfield  {author} {\bibinfo {author} {\bibfnamefont {Z.}~\bibnamefont
  {Li}}, \bibinfo {author} {\bibfnamefont {K.~G.}\ \bibnamefont {Pradeep}},
  \bibinfo {author} {\bibfnamefont {Y.}~\bibnamefont {Deng}}, \bibinfo {author}
  {\bibfnamefont {D.}~\bibnamefont {Raabe}}, \ and\ \bibinfo {author}
  {\bibfnamefont {C.~C.}\ \bibnamefont {Tasan}},\ }\href@noop {} {\bibfield
  {journal} {\bibinfo  {journal} {Nature}\ }\textbf {\bibinfo {volume} {534}},\
  \bibinfo {pages} {227} (\bibinfo {year} {2016})}\BibitemShut {NoStop}%
\bibitem [{\citenamefont {Murty}\ \emph {et~al.}(2014)\citenamefont {Murty},
  \citenamefont {Yeh},\ and\ \citenamefont {Ranganathan}}]{MYW14}%
  \BibitemOpen
  \bibfield  {author} {\bibinfo {author} {\bibfnamefont {B.~S.}\ \bibnamefont
  {Murty}}, \bibinfo {author} {\bibfnamefont {J.~W.}\ \bibnamefont {Yeh}}, \
  and\ \bibinfo {author} {\bibfnamefont {S.}~\bibnamefont {Ranganathan}},\
  }\href {\doibase 10.1016/b978-0-12-800251-3.00011-0} {\emph {\bibinfo {title}
  {High-entropy alloys}}}\ (\bibinfo  {publisher} {Butterworth-Heinemann},\
  \bibinfo {address} {London},\ \bibinfo {year} {2014})\BibitemShut {NoStop}%
\bibitem [{\citenamefont {Gao}\ \emph {et~al.}(2016)\citenamefont {Gao},
  \citenamefont {Yeh}, \citenamefont {Liaw},\ and\ \citenamefont
  {Zhang}}]{GYW16}%
  \BibitemOpen
  \bibfield  {author} {\bibinfo {author} {\bibfnamefont {M.~C.}\ \bibnamefont
  {Gao}}, \bibinfo {author} {\bibfnamefont {J.-W.}\ \bibnamefont {Yeh}},
  \bibinfo {author} {\bibfnamefont {P.~K.}\ \bibnamefont {Liaw}}, \ and\
  \bibinfo {author} {\bibfnamefont {Y.}~\bibnamefont {Zhang}},\ }\href@noop {}
  {\emph {\bibinfo {title} {High-entropy alloys: Fundamentals and
  applications}}}\ (\bibinfo  {publisher} {Springer},\ \bibinfo {address}
  {International Publishing Switzerland},\ \bibinfo {year} {2016})\BibitemShut
  {NoStop}%
\bibitem{IGK2018}
Y. Ikeda, B. Grabowski, F. K{\"o}rmann, Mater. Charact. {\bf 147} , 464 (2019).
\bibitem [{\citenamefont {Engin}\ \emph {et~al.}(2008)\citenamefont {Engin},
  \citenamefont {Sandoval},\ and\ \citenamefont
  {Urbassek}}]{engin2008characterization}%
  \BibitemOpen
  \bibfield  {author} {\bibinfo {author} {\bibfnamefont {C.}~\bibnamefont
  {Engin}}, \bibinfo {author} {\bibfnamefont {L.}~\bibnamefont {Sandoval}}, \
  and\ \bibinfo {author} {\bibfnamefont {H.~M.}\ \bibnamefont {Urbassek}},\
  }\href@noop {} {\bibfield  {journal} {\bibinfo  {journal} {Model. Simul. Mater. Sci. Eng}\ }\textbf {\bibinfo {volume}
  {16}},\ \bibinfo {pages} {035005} (\bibinfo {year} {2008})}\BibitemShut
  {NoStop}%
\bibitem [{\citenamefont {Grabowski}\ \emph {et~al.}(2011)\citenamefont
  {Grabowski}, \citenamefont {S\"oderlind}, \citenamefont {Hickel},\ and\
  \citenamefont {Neugebauer}}]{GSH2011}%
  \BibitemOpen
  \bibfield  {author} {\bibinfo {author} {\bibfnamefont {B.}~\bibnamefont
  {Grabowski}}, \bibinfo {author} {\bibfnamefont {P.}~\bibnamefont
  {S\"oderlind}}, \bibinfo {author} {\bibfnamefont {T.}~\bibnamefont {Hickel}},
  \ and\ \bibinfo {author} {\bibfnamefont {J.}~\bibnamefont {Neugebauer}},\
  }\href@noop {} {\bibfield  {journal} {\bibinfo  {journal} {Phys. Rev. B}\
  }\textbf {\bibinfo {volume} {84}},\ \bibinfo {pages} {214107} (\bibinfo
  {year} {2011})}\BibitemShut {NoStop}%
\bibitem [{\citenamefont {Zhu}\ \emph {et~al.}(2017)\citenamefont {Zhu},
  \citenamefont {Grabowski},\ and\ \citenamefont {Neugebauer}}]{ZGN2017}%
  \BibitemOpen
  \bibfield  {author} {\bibinfo {author} {\bibfnamefont {L.-F.}\ \bibnamefont
  {Zhu}}, \bibinfo {author} {\bibfnamefont {B.}~\bibnamefont {Grabowski}}, \
  and\ \bibinfo {author} {\bibfnamefont {J.}~\bibnamefont {Neugebauer}},\
  }\href@noop {} {\bibfield  {journal} {\bibinfo  {journal} {Phys. Rev. B}\
  }\textbf {\bibinfo {volume} {96}},\ \bibinfo {pages} {224202} (\bibinfo
  {year} {2017})}\BibitemShut {NoStop}%
\bibitem [{\citenamefont {Duff}\ \emph
  {et~al.}(2015{\natexlab{a}})\citenamefont {Duff}, \citenamefont {Davey},
  \citenamefont {Korbmacher}, \citenamefont {Glensk}, \citenamefont
  {Grabowski}, \citenamefont {Neugebauer},\ and\ \citenamefont
  {Finnis}}]{Duf2015}%
  \BibitemOpen
  \bibfield  {author} {\bibinfo {author} {\bibfnamefont {A.~I.}\ \bibnamefont
  {Duff}}, \bibinfo {author} {\bibfnamefont {T.}~\bibnamefont {Davey}},
  \bibinfo {author} {\bibfnamefont {D.}~\bibnamefont {Korbmacher}}, \bibinfo
  {author} {\bibfnamefont {A.}~\bibnamefont {Glensk}}, \bibinfo {author}
  {\bibfnamefont {B.}~\bibnamefont {Grabowski}}, \bibinfo {author}
  {\bibfnamefont {J.}~\bibnamefont {Neugebauer}}, \ and\ \bibinfo {author}
  {\bibfnamefont {M.~W.}\ \bibnamefont {Finnis}},\ }\href {\doibase
  10.1103/PhysRevB.91.214311} {\bibfield  {journal} {\bibinfo  {journal} {Phys.
  Rev. B}\ }\textbf {\bibinfo {volume} {91}},\ \bibinfo {pages} {214311}
  (\bibinfo {year} {2015}{\natexlab{a}})}\BibitemShut {NoStop}%
\bibitem [{\citenamefont {Glensk}\ \emph {et~al.}(2015)\citenamefont {Glensk},
  \citenamefont {Grabowski}, \citenamefont {Hickel},\ and\ \citenamefont
  {Neugebauer}}]{GGH2015}%
  \BibitemOpen
  \bibfield  {author} {\bibinfo {author} {\bibfnamefont {A.}~\bibnamefont
  {Glensk}}, \bibinfo {author} {\bibfnamefont {B.}~\bibnamefont {Grabowski}},
  \bibinfo {author} {\bibfnamefont {T.}~\bibnamefont {Hickel}}, \ and\ \bibinfo
  {author} {\bibfnamefont {J.}~\bibnamefont {Neugebauer}},\ }\href@noop {}
  {\bibfield  {journal} {\bibinfo  {journal} {Phys. Rev. Lett.}\ }\textbf
  {\bibinfo {volume} {114}},\ \bibinfo {pages} {195901} (\bibinfo {year}
  {2015})}\BibitemShut {NoStop}%
\bibitem [{\citenamefont {Senkov}\ \emph {et~al.}(2011)\citenamefont {Senkov},
  \citenamefont {Wilks}, \citenamefont {Scott},\ and\ \citenamefont
  {Miracle}}]{SENKOV2011698}%
  \BibitemOpen
  \bibfield  {author} {\bibinfo {author} {\bibfnamefont {O.}~\bibnamefont
  {Senkov}}, \bibinfo {author} {\bibfnamefont {G.}~\bibnamefont {Wilks}},
  \bibinfo {author} {\bibfnamefont {J.}~\bibnamefont {Scott}}, \ and\ \bibinfo
  {author} {\bibfnamefont {D.}~\bibnamefont {Miracle}},\ }\href {\doibase
  https://doi.org/10.1016/j.intermet.2011.01.004} {\bibfield  {journal}
  {\bibinfo  {journal} {Intermetallics}\ }\textbf {\bibinfo {volume} {19}},\
  \bibinfo {pages} {698 } (\bibinfo {year} {2011})}\BibitemShut {NoStop}%
\bibitem [{\citenamefont {Senkov}\ \emph {et~al.}(2010)\citenamefont {Senkov},
  \citenamefont {Wilks}, \citenamefont {Miracle}, \citenamefont {Chuang},\ and\
  \citenamefont {Liaw}}]{SENKOV20101758}%
  \BibitemOpen
  \bibfield  {author} {\bibinfo {author} {\bibfnamefont {O.}~\bibnamefont
  {Senkov}}, \bibinfo {author} {\bibfnamefont {G.}~\bibnamefont {Wilks}},
  \bibinfo {author} {\bibfnamefont {D.}~\bibnamefont {Miracle}}, \bibinfo
  {author} {\bibfnamefont {C.}~\bibnamefont {Chuang}}, \ and\ \bibinfo {author}
  {\bibfnamefont {P.}~\bibnamefont {Liaw}},\ }\href {\doibase
  https://doi.org/10.1016/j.intermet.2010.05.014} {\bibfield  {journal}
  {\bibinfo  {journal} {Intermetallics}\ }\textbf {\bibinfo {volume} {18}},\
  \bibinfo {pages} {1758 } (\bibinfo {year} {2010})}\BibitemShut {NoStop}%
\bibitem [{\citenamefont {Zhang}\ \emph {et~al.}(2018)\citenamefont {Zhang},
  \citenamefont {Grabowski}, \citenamefont {Hickel},\ and\ \citenamefont
  {Neugebauer}}]{ZGH2018}%
  \BibitemOpen
  \bibfield  {author} {\bibinfo {author} {\bibfnamefont {X.}~\bibnamefont
  {Zhang}}, \bibinfo {author} {\bibfnamefont {B.}~\bibnamefont {Grabowski}},
  \bibinfo {author} {\bibfnamefont {T.}~\bibnamefont {Hickel}}, \ and\ \bibinfo
  {author} {\bibfnamefont {J.}~\bibnamefont {Neugebauer}},\ }\href@noop {}
  {\bibfield  {journal} {\bibinfo  {journal} {Comp. Mater. Sci.}\ }\textbf
  {\bibinfo {volume} {148}},\ \bibinfo {pages} {249} (\bibinfo {year}
  {2018})}\BibitemShut {NoStop}%
\bibitem [{\citenamefont {K\"ormann}\ \emph {et~al.}(2017)\citenamefont
  {K\"ormann}, \citenamefont {Ikeda}, \citenamefont {Grabowski},\ and\
  \citenamefont {Sluiter}}]{KIG17}%
  \BibitemOpen
  \bibfield  {author} {\bibinfo {author} {\bibfnamefont {F.}~\bibnamefont
  {K\"ormann}}, \bibinfo {author} {\bibfnamefont {Y.}~\bibnamefont {Ikeda}},
  \bibinfo {author} {\bibfnamefont {B.}~\bibnamefont {Grabowski}}, \ and\
  \bibinfo {author} {\bibfnamefont {M.~H.~F.}\ \bibnamefont {Sluiter}},\ }\href
  {https://doi.org/10.1038/s41524-017-0037-8} {\bibfield  {journal} {\bibinfo
  {journal} {npj Comput. Mater.}\ }\textbf {\bibinfo {volume} {3}},\ \bibinfo
  {pages} {36} (\bibinfo {year} {2017})}\BibitemShut {NoStop}%
\bibitem [{\citenamefont {Souvatzis}\ \emph {et~al.}(2008)\citenamefont
  {Souvatzis}, \citenamefont {Eriksson}, \citenamefont {Katsnelson},\ and\
  \citenamefont {Rudin}}]{Sou2008}%
  \BibitemOpen
  \bibfield  {author} {\bibinfo {author} {\bibfnamefont {P.}~\bibnamefont
  {Souvatzis}}, \bibinfo {author} {\bibfnamefont {O.}~\bibnamefont {Eriksson}},
  \bibinfo {author} {\bibfnamefont {M.~I.}\ \bibnamefont {Katsnelson}}, \ and\
  \bibinfo {author} {\bibfnamefont {S.~P.}\ \bibnamefont {Rudin}},\ }\href
  {\doibase 10.1103/PhysRevLett.100.095901} {\bibfield  {journal} {\bibinfo
  {journal} {Phys. Rev. Lett.}\ }\textbf {\bibinfo {volume} {100}},\ \bibinfo
  {pages} {095901} (\bibinfo {year} {2008})}\BibitemShut {NoStop}%
\bibitem [{\citenamefont {Souvatzis}\ \emph {et~al.}(2009)\citenamefont
  {Souvatzis}, \citenamefont {Eriksson}, \citenamefont {Katsnelson},\ and\
  \citenamefont {Rudin}}]{Sou2009}%
  \BibitemOpen
  \bibfield  {author} {\bibinfo {author} {\bibfnamefont {P.}~\bibnamefont
  {Souvatzis}}, \bibinfo {author} {\bibfnamefont {O.}~\bibnamefont {Eriksson}},
  \bibinfo {author} {\bibfnamefont {M.}~\bibnamefont {Katsnelson}}, \ and\
  \bibinfo {author} {\bibfnamefont {S.}~\bibnamefont {Rudin}},\ }\href
  {\doibase https://doi.org/10.1016/j.commatsci.2008.06.016} {\bibfield
  {journal} {\bibinfo  {journal} {Comp. Mater. Sci.}\ }\textbf
  {\bibinfo {volume} {44}},\ \bibinfo {pages} {888 } (\bibinfo {year}
  {2009})}\BibitemShut {NoStop}%
\bibitem [{\citenamefont {Hellman}\ \emph {et~al.}(2013)\citenamefont
  {Hellman}, \citenamefont {Steneteg}, \citenamefont {Abrikosov},\ and\
  \citenamefont {Simak}}]{HSA2013}%
  \BibitemOpen
  \bibfield  {author} {\bibinfo {author} {\bibfnamefont {O.}~\bibnamefont
  {Hellman}}, \bibinfo {author} {\bibfnamefont {P.}~\bibnamefont {Steneteg}},
  \bibinfo {author} {\bibfnamefont {I.~A.}\ \bibnamefont {Abrikosov}}, \ and\
  \bibinfo {author} {\bibfnamefont {S.~I.}~\bibnamefont {Simak}},\ }\href@noop {}
  {\bibfield  {journal} {\bibinfo  {journal} {Phys. Rev. B}\ }\textbf {\bibinfo
  {volume} {87}},\ \bibinfo {pages} {104111} (\bibinfo {year}
  {2013})}\BibitemShut {NoStop}%
\bibitem [{\citenamefont {Errea}\ \emph {et~al.}(2014)\citenamefont {Errea},
  \citenamefont {Calandra},\ and\ \citenamefont {Mauri}}]{PhysRevB.89.064302}%
  \BibitemOpen
  \bibfield  {author} {\bibinfo {author} {\bibfnamefont {I.}~\bibnamefont
  {Errea}}, \bibinfo {author} {\bibfnamefont {M.}~\bibnamefont {Calandra}}, \
  and\ \bibinfo {author} {\bibfnamefont {F.}~\bibnamefont {Mauri}},\ }\href
  {\doibase 10.1103/PhysRevB.89.064302} {\bibfield  {journal} {\bibinfo
  {journal} {Phys. Rev. B}\ }\textbf {\bibinfo {volume} {89}},\ \bibinfo
  {pages} {064302} (\bibinfo {year} {2014})}\BibitemShut {NoStop}%
\bibitem [{\citenamefont {Sun}\ \emph {et~al.}(2014)\citenamefont {Sun},
  \citenamefont {Zhang},\ and\ \citenamefont
  {Wentzcovitch}}]{PhysRevB.89.094109}%
  \BibitemOpen
  \bibfield  {author} {\bibinfo {author} {\bibfnamefont {T.}~\bibnamefont
  {Sun}}, \bibinfo {author} {\bibfnamefont {D.-B.}\ \bibnamefont {Zhang}}, \
  and\ \bibinfo {author} {\bibfnamefont {R.~M.}\ \bibnamefont {Wentzcovitch}},\
  }\href {\doibase 10.1103/PhysRevB.89.094109} {\bibfield  {journal} {\bibinfo
  {journal} {Phys. Rev. B}\ }\textbf {\bibinfo {volume} {89}},\ \bibinfo
  {pages} {094109} (\bibinfo {year} {2014})}\BibitemShut {NoStop}%
\bibitem [{\citenamefont {Carreras}\ \emph {et~al.}(2017)\citenamefont
  {Carreras}, \citenamefont {Togo},\ and\ \citenamefont
  {Tanaka}}]{CARRERAS2017221}%
  \BibitemOpen
  \bibfield  {author} {\bibinfo {author} {\bibfnamefont {A.}~\bibnamefont
  {Carreras}}, \bibinfo {author} {\bibfnamefont {A.}~\bibnamefont {Togo}}, \
  and\ \bibinfo {author} {\bibfnamefont {I.}~\bibnamefont {Tanaka}},\ }\href
  {\doibase https://doi.org/10.1016/j.cpc.2017.08.017} {\bibfield  {journal}
  {\bibinfo  {journal} {Computer Physics Communications}\ }\textbf {\bibinfo
  {volume} {221}},\ \bibinfo {pages} {221 } (\bibinfo {year}
  {2017})}\BibitemShut {NoStop}%
\bibitem [{\citenamefont {Alf\`e}\ \emph {et~al.}(2001)\citenamefont {Alf\`e},
  \citenamefont {Price},\ and\ \citenamefont {Gillan}}]{Alf2001}%
  \BibitemOpen
  \bibfield  {author} {\bibinfo {author} {\bibfnamefont {D.}~\bibnamefont
  {Alf\`e}}, \bibinfo {author} {\bibfnamefont {G.~D.}\ \bibnamefont {Price}}, \
  and\ \bibinfo {author} {\bibfnamefont {M.~J.}\ \bibnamefont {Gillan}},\
  }\href {\doibase 10.1103/PhysRevB.64.045123} {\bibfield  {journal} {\bibinfo
  {journal} {Phys. Rev. B}\ }\textbf {\bibinfo {volume} {64}},\ \bibinfo
  {pages} {045123} (\bibinfo {year} {2001})}\BibitemShut {NoStop}%
\bibitem [{\citenamefont {Alf\`e}\ \emph
  {et~al.}(2002{\natexlab{a}})\citenamefont {Alf\`e}, \citenamefont {Gillan},\
  and\ \citenamefont {Price}}]{Alf2002-1}%
  \BibitemOpen
  \bibfield  {author} {\bibinfo {author} {\bibfnamefont {D.}~\bibnamefont
  {Alf\`e}}, \bibinfo {author} {\bibfnamefont {M.~J.}\ \bibnamefont {Gillan}},
  \ and\ \bibinfo {author} {\bibfnamefont {G.~D.}\ \bibnamefont {Price}},\
  }\href {\doibase 10.1063/1.1460865} {\bibfield  {journal} {\bibinfo
  {journal} {J. Chem. Phys.}\ }\textbf {\bibinfo {volume}
  {116}},\ \bibinfo {pages} {6170} (\bibinfo {year} {2002}{\natexlab{a}})},\
  \Eprint {http://arxiv.org/abs/https://doi.org/10.1063/1.1460865}
  {https://doi.org/10.1063/1.1460865} \BibitemShut {NoStop}%
\bibitem [{\citenamefont {Alf\`e}\ \emph
  {et~al.}(2002{\natexlab{b}})\citenamefont {Alf\`e}, \citenamefont {Price},\
  and\ \citenamefont {Gillan}}]{Alf2002-2}%
  \BibitemOpen
  \bibfield  {author} {\bibinfo {author} {\bibfnamefont {D.}~\bibnamefont
  {Alf\`e}}, \bibinfo {author} {\bibfnamefont {G.~D.}\ \bibnamefont {Price}}, \
  and\ \bibinfo {author} {\bibfnamefont {M.~J.}\ \bibnamefont {Gillan}},\
  }\href {\doibase 10.1103/PhysRevB.65.165118} {\bibfield  {journal} {\bibinfo
  {journal} {Phys. Rev. B}\ }\textbf {\bibinfo {volume} {65}},\ \bibinfo
  {pages} {165118} (\bibinfo {year} {2002}{\natexlab{b}})}\BibitemShut
  {NoStop}%
\bibitem [{\citenamefont {Vocadlo}\ \emph {et~al.}(2003)\citenamefont
  {Vocadlo}, \citenamefont {Alf\`e}, \citenamefont {Gillan}, \citenamefont
  {Wood}, \citenamefont {Brodholt},\ and\ \citenamefont {Price}}]{Lid2003}%
  \BibitemOpen
  \bibfield  {author} {\bibinfo {author} {\bibfnamefont {L.}~\bibnamefont
  {Vocadlo}}, \bibinfo {author} {\bibfnamefont {D.}~\bibnamefont {Alf\`e}},
  \bibinfo {author} {\bibfnamefont {M.~J.}\ \bibnamefont {Gillan}}, \bibinfo
  {author} {\bibfnamefont {I.~G.}\ \bibnamefont {Wood}}, \bibinfo {author}
  {\bibfnamefont {J.~P.}\ \bibnamefont {Brodholt}}, \ and\ \bibinfo {author}
  {\bibfnamefont {G.~D.}\ \bibnamefont {Price}},\ }\href@noop {} {\bibfield
  {journal} {\bibinfo  {journal} {Nature}\ }\textbf {\bibinfo {volume} {424}},\
  \bibinfo {pages} {536} (\bibinfo {year} {2003})}\BibitemShut {NoStop}%
\bibitem [{\citenamefont {Moustafa}\ \emph {et~al.}(2017)\citenamefont
  {Moustafa}, \citenamefont {Schultz}, \citenamefont {Zurek},\ and\
  \citenamefont {Kofke}}]{Mou2017}%
  \BibitemOpen
  \bibfield  {author} {\bibinfo {author} {\bibfnamefont {S.~G.}\ \bibnamefont
  {Moustafa}}, \bibinfo {author} {\bibfnamefont {A.~J.}\ \bibnamefont
  {Schultz}}, \bibinfo {author} {\bibfnamefont {E.}~\bibnamefont {Zurek}}, \
  and\ \bibinfo {author} {\bibfnamefont {D.~A.}\ \bibnamefont {Kofke}},\ }\href
  {\doibase 10.1103/PhysRevB.96.014117} {\bibfield  {journal} {\bibinfo
  {journal} {Phys. Rev. B}\ }\textbf {\bibinfo {volume} {96}},\ \bibinfo
  {pages} {014117} (\bibinfo {year} {2017})}\BibitemShut {NoStop}%
\bibitem [{sup()}]{supp}%
  \BibitemOpen
  \href@noop {} {}\bibinfo {note} {See Supplementary Information.}\BibitemShut
  {Stop}%
\bibitem [{\citenamefont {{Podryabinkin}}\ and\ \citenamefont
  {{Shapeev}}(2016)}]{podryabinkin2016-active-learning}%
  \BibitemOpen
  \bibfield  {author} {\bibinfo {author} {\bibfnamefont {E.~V.}\ \bibnamefont
  {{Podryabinkin}}}\ and\ \bibinfo {author} {\bibfnamefont {A.~V.}\
  \bibnamefont {{Shapeev}}},\ }\href@noop {} {\bibfield  {journal} {\bibinfo
  {journal} {ArXiv e-prints}\ } (\bibinfo {year} {2016})},\ \Eprint
  {http://arxiv.org/abs/1611.09346} {arXiv:1611.09346 [physics.comp-ph]}
  \BibitemShut {NoStop}%
\bibitem [{\citenamefont {{Rupp}}\ \emph {et~al.}(2018)\citenamefont {{Rupp}},
  \citenamefont {{von Lilienfeld}},\ and\ \citenamefont
  {{Burke}}}]{rupp2018-ml-chemistry-editorial}%
  \BibitemOpen
  \bibfield  {author} {\bibinfo {author} {\bibfnamefont {M.}~\bibnamefont
  {{Rupp}}}, \bibinfo {author} {\bibfnamefont {O.~A.}\ \bibnamefont {{von
  Lilienfeld}}}, \ and\ \bibinfo {author} {\bibfnamefont {K.}~\bibnamefont
  {{Burke}}},\ }\href {\doibase 10.1063/1.5043213} {\bibfield  {journal}
  {\bibinfo  {journal} {\jcp}\ }\textbf {\bibinfo {volume} {148}},\ \bibinfo
  {eid} {241401} (\bibinfo {year} {2018})}\BibitemShut {NoStop}%
\bibitem [{\citenamefont {Behler}\ and\ \citenamefont
  {Parrinello}(2007)}]{PhysRevLett.98.146401}%
  \BibitemOpen
  \bibfield  {author} {\bibinfo {author} {\bibfnamefont {J.}~\bibnamefont
  {Behler}}\ and\ \bibinfo {author} {\bibfnamefont {M.}~\bibnamefont
  {Parrinello}},\ }\href {\doibase 10.1103/PhysRevLett.98.146401} {\bibfield
  {journal} {\bibinfo  {journal} {Phys. Rev. Lett.}\ }\textbf {\bibinfo
  {volume} {98}},\ \bibinfo {pages} {146401} (\bibinfo {year}
  {2007})}\BibitemShut {NoStop}%
\bibitem [{\citenamefont {Behler}(2011)}]{doi:10.1063/1.3553717}%
  \BibitemOpen
  \bibfield  {author} {\bibinfo {author} {\bibfnamefont {J.}~\bibnamefont
  {Behler}},\ }\href {\doibase 10.1063/1.3553717} {\bibfield  {journal}
  {\bibinfo  {journal} {J. Chem. Phys.}\ }\textbf {\bibinfo
  {volume} {134}},\ \bibinfo {pages} {074106} (\bibinfo {year}
  {2011})}\BibitemShut {NoStop}%
\bibitem [{\citenamefont {Bart\'ok}\ \emph {et~al.}(2010)\citenamefont
  {Bart\'ok}, \citenamefont {Payne}, \citenamefont {Kondor},\ and\
  \citenamefont {Cs\'anyi}}]{PhysRevLett.104.136403}%
  \BibitemOpen
  \bibfield  {author} {\bibinfo {author} {\bibfnamefont {A.~P.}\ \bibnamefont
  {Bart\'ok}}, \bibinfo {author} {\bibfnamefont {M.~C.}\ \bibnamefont {Payne}},
  \bibinfo {author} {\bibfnamefont {R.}~\bibnamefont {Kondor}}, \ and\ \bibinfo
  {author} {\bibfnamefont {G.}~\bibnamefont {Cs\'anyi}},\ }\href {\doibase
  10.1103/PhysRevLett.104.136403} {\bibfield  {journal} {\bibinfo  {journal}
  {Phys. Rev. Lett.}\ }\textbf {\bibinfo {volume} {104}},\ \bibinfo {pages}
  {136403} (\bibinfo {year} {2010})}\BibitemShut {NoStop}%
\bibitem [{\citenamefont {Bart{\'o}k}\ \emph {et~al.}(2017)\citenamefont
  {Bart{\'o}k}, \citenamefont {De}, \citenamefont {Poelking}, \citenamefont
  {Bernstein}, \citenamefont {Kermode}, \citenamefont {Cs{\'a}nyi},\ and\
  \citenamefont {Ceriotti}}]{Bartoke1701816}%
  \BibitemOpen
  \bibfield  {author} {\bibinfo {author} {\bibfnamefont {A.~P.}\ \bibnamefont
  {Bart{\'o}k}}, \bibinfo {author} {\bibfnamefont {S.}~\bibnamefont {De}},
  \bibinfo {author} {\bibfnamefont {C.}~\bibnamefont {Poelking}}, \bibinfo
  {author} {\bibfnamefont {N.}~\bibnamefont {Bernstein}}, \bibinfo {author}
  {\bibfnamefont {J.~R.}\ \bibnamefont {Kermode}}, \bibinfo {author}
  {\bibfnamefont {G.}~\bibnamefont {Cs{\'a}nyi}}, \ and\ \bibinfo {author}
  {\bibfnamefont {M.}~\bibnamefont {Ceriotti}},\ }\href
  {http://advances.sciencemag.org/content/3/12/e1701816} {\bibfield  {journal}
  {\bibinfo  {journal} {Sci. Adv.}\ }\textbf {\bibinfo {volume} {3}}
  (\bibinfo {year} {2017})}\BibitemShut {NoStop}%
\bibitem [{\citenamefont {Seko}\ \emph {et~al.}(2015)\citenamefont {Seko},
  \citenamefont {Takahashi},\ and\ \citenamefont
  {Tanaka}}]{PhysRevB.92.054113}%
  \BibitemOpen
  \bibfield  {author} {\bibinfo {author} {\bibfnamefont {A.}~\bibnamefont
  {Seko}}, \bibinfo {author} {\bibfnamefont {A.}~\bibnamefont {Takahashi}}, \
  and\ \bibinfo {author} {\bibfnamefont {I.}~\bibnamefont {Tanaka}},\ }\href
  {\doibase 10.1103/PhysRevB.92.054113} {\bibfield  {journal} {\bibinfo
  {journal} {Phys. Rev. B}\ }\textbf {\bibinfo {volume} {92}},\ \bibinfo
  {pages} {054113} (\bibinfo {year} {2015})}\BibitemShut {NoStop}%
\bibitem [{\citenamefont {Takahashi}\ \emph {et~al.}(2017)\citenamefont
  {Takahashi}, \citenamefont {Seko},\ and\ \citenamefont
  {Tanaka}}]{PhysRevMaterials.1.063801}%
  \BibitemOpen
  \bibfield  {author} {\bibinfo {author} {\bibfnamefont {A.}~\bibnamefont
  {Takahashi}}, \bibinfo {author} {\bibfnamefont {A.}~\bibnamefont {Seko}}, \
  and\ \bibinfo {author} {\bibfnamefont {I.}~\bibnamefont {Tanaka}},\ }\href
  {\doibase 10.1103/PhysRevMaterials.1.063801} {\bibfield  {journal} {\bibinfo
  {journal} {Phys. Rev. Materials}\ }\textbf {\bibinfo {volume} {1}},\ \bibinfo
  {pages} {063801} (\bibinfo {year} {2017})}\BibitemShut {NoStop}%
\bibitem [{\citenamefont {Stecher}\ \emph {et~al.}(2014)\citenamefont
  {Stecher}, \citenamefont {Bernstein},\ and\ \citenamefont
  {Cs{\'a}nyi}}]{stecher2014-free-energy-peptides}%
  \BibitemOpen
  \bibfield  {author} {\bibinfo {author} {\bibfnamefont {T.}~\bibnamefont
  {Stecher}}, \bibinfo {author} {\bibfnamefont {N.}~\bibnamefont {Bernstein}},
  \ and\ \bibinfo {author} {\bibfnamefont {G.}~\bibnamefont {Cs{\'a}nyi}},\
  }\href@noop {} {\bibfield  {journal} {\bibinfo  {journal} {J. Chem. Theory Comput.}\ }\textbf {\bibinfo {volume} {10}},\
  \bibinfo {pages} {4079} (\bibinfo {year} {2014})}\BibitemShut {NoStop}%
\bibitem [{\citenamefont {{Gubaev}}\ \emph {et~al.}(2018)\citenamefont
  {{Gubaev}}, \citenamefont {{Podryabinkin}}, \citenamefont {{Hart}},\ and\
  \citenamefont {{Shapeev}}}]{2018arXiv180610567G}%
  \BibitemOpen
  \bibfield  {author} {\bibinfo {author} {\bibfnamefont {K.}~\bibnamefont
  {{Gubaev}}}, \bibinfo {author} {\bibfnamefont {E.~V.}\ \bibnamefont
  {{Podryabinkin}}}, \bibinfo {author} {\bibfnamefont {G.~L.~W.}\ \bibnamefont
  {{Hart}}}, \ and\ \bibinfo {author} {\bibfnamefont {A.~V.}\ \bibnamefont
  {{Shapeev}}},\ }\href@noop {} {\bibfield  {journal} {\bibinfo  {journal}
  {ArXiv e-prints}\ } (\bibinfo {year} {2018})},\ \Eprint
  {http://arxiv.org/abs/1806.10567} {arXiv:1806.10567 [cond-mat.mtrl-sci]}
  \BibitemShut {NoStop}%
\bibitem [{\citenamefont {Shapeev}(2016)}]{shapeev2016mtp}%
  \BibitemOpen
  \bibfield  {author} {\bibinfo {author} {\bibfnamefont {A.~V.}\ \bibnamefont
  {Shapeev}},\ }\href@noop {} {\bibfield  {journal} {\bibinfo  {journal}
  {Multiscale Model. Simul.}\ }\textbf {\bibinfo {volume} {14}},\
  \bibinfo {pages} {1153} (\bibinfo {year} {2016})}\BibitemShut {NoStop}%
\bibitem {comp}
C. Nyshadham, M. Rupp, B. Bekker, A. V. Shapeev, T. Mueller, C. W. Rosenbrock, G. Cs{\'a}nyi, D. W. Wingate and G. L. Hart, ArXiv e-prints (2018) arXiv:1809.09203.
\bibitem [{\citenamefont {Zunger}\ \emph {et~al.}(1990)\citenamefont {Zunger},
  \citenamefont {Wei}, \citenamefont {Ferreira},\ and\ \citenamefont
  {Bernard}}]{Zunger_PRL_1990_Special}%
  \BibitemOpen
  \bibfield  {author} {\bibinfo {author} {\bibfnamefont {A.}~\bibnamefont
  {Zunger}}, \bibinfo {author} {\bibfnamefont {S.-H.}\ \bibnamefont {Wei}},
  \bibinfo {author} {\bibfnamefont {L.~G.}\ \bibnamefont {Ferreira}}, \ and\
  \bibinfo {author} {\bibfnamefont {J.~E.}\ \bibnamefont {Bernard}},\ }\href
  {\doibase 10.1103/PhysRevLett.65.353} {\bibfield  {journal} {\bibinfo
  {journal} {Phys. Rev. Lett.}\ }\textbf {\bibinfo {volume} {65}},\ \bibinfo
  {pages} {353} (\bibinfo {year} {1990})}\BibitemShut {NoStop}%
\bibitem [{\citenamefont {Kresse}\ and\ \citenamefont
  {Furthm\"uller}(1996{\natexlab{a}})}]{KF1996}%
  \BibitemOpen
  \bibfield  {author} {\bibinfo {author} {\bibfnamefont {G.}~\bibnamefont
  {Kresse}}\ and\ \bibinfo {author} {\bibfnamefont {J.}~\bibnamefont
  {Furthm\"uller}},\ }\href@noop {} {\bibfield  {journal} {\bibinfo  {journal}
  {Comp. Mater. Sci.}\ }\textbf {\bibinfo {volume} {6}},\ \bibinfo {pages} {15}
  (\bibinfo {year} {1996}{\natexlab{a}})}\BibitemShut {NoStop}%
\bibitem [{\citenamefont {Kresse}\ and\ \citenamefont
  {Furthm\"uller}(1996{\natexlab{b}})}]{KF1996b}%
  \BibitemOpen
  \bibfield  {author} {\bibinfo {author} {\bibfnamefont {G.}~\bibnamefont
  {Kresse}}\ and\ \bibinfo {author} {\bibfnamefont {J.}~\bibnamefont
  {Furthm\"uller}},\ }\href@noop {} {\bibfield  {journal} {\bibinfo  {journal}
  {Phys. Rev. B}\ }\textbf {\bibinfo {volume} {54}},\ \bibinfo {pages} {11169}
  (\bibinfo {year} {1996}{\natexlab{b}})}\BibitemShut {NoStop}%
\bibitem [{\citenamefont {Bl\"ochl}(1994)}]{Blo1994}%
  \BibitemOpen
  \bibfield  {author} {\bibinfo {author} {\bibfnamefont {P.~E.}\ \bibnamefont
  {Bl\"ochl}},\ }\href@noop {} {\bibfield  {journal} {\bibinfo  {journal}
  {Phys. Rev. B}\ }\textbf {\bibinfo {volume} {50}},\ \bibinfo {pages} {17953}
  (\bibinfo {year} {1994})}\BibitemShut {NoStop}%
\bibitem [{\citenamefont {Perdew}\ \emph {et~al.}(1996)\citenamefont {Perdew},
  \citenamefont {Burke},\ and\ \citenamefont {Ernzerhof}}]{PBE1996}%
  \BibitemOpen
  \bibfield  {author} {\bibinfo {author} {\bibfnamefont {J.~P.}\ \bibnamefont
  {Perdew}}, \bibinfo {author} {\bibfnamefont {K.}~\bibnamefont {Burke}}, \
  and\ \bibinfo {author} {\bibfnamefont {M.}~\bibnamefont {Ernzerhof}},\
  }\href@noop {} {\bibfield  {journal} {\bibinfo  {journal} {Phys. Rev. Lett.}\
  }\textbf {\bibinfo {volume} {77}},\ \bibinfo {pages} {3865} (\bibinfo {year}
  {1996})}\BibitemShut {NoStop}%
\bibitem [{\citenamefont {Zhang}\ \emph {et~al.}(2017)\citenamefont {Zhang},
  \citenamefont {Grabowski}, \citenamefont {K\"ormann}, \citenamefont
  {Freysoldt},\ and\ \citenamefont {Neugebauer}}]{Zhang2017}%
  \BibitemOpen
  \bibfield  {author} {\bibinfo {author} {\bibfnamefont {X.}~\bibnamefont
  {Zhang}}, \bibinfo {author} {\bibfnamefont {B.}~\bibnamefont {Grabowski}},
  \bibinfo {author} {\bibfnamefont {F.}~\bibnamefont {K\"ormann}}, \bibinfo
  {author} {\bibfnamefont {C.}~\bibnamefont {Freysoldt}}, \ and\ \bibinfo
  {author} {\bibfnamefont {J.}~\bibnamefont {Neugebauer}},\ }\href@noop {}
  {\bibfield  {journal} {\bibinfo  {journal} {Phys. Rev. B}\ }\textbf
  {\bibinfo {volume} {95}},\ \bibinfo {pages} {165126} (\bibinfo {year}
  {2017})}\BibitemShut {NoStop}%
\bibitem [{\citenamefont {Gong}\ \emph {et~al.}(2018)\citenamefont {Gong},
  \citenamefont {Grabowski}, \citenamefont {Glensk}, \citenamefont {K\"ormann},
  \citenamefont {Neugebauer},\ and\ \citenamefont {Reed}}]{GGG2018}%
  \BibitemOpen
  \bibfield  {author} {\bibinfo {author} {\bibfnamefont {Y.}~\bibnamefont
  {Gong}}, \bibinfo {author} {\bibfnamefont {B.}~\bibnamefont {Grabowski}},
  \bibinfo {author} {\bibfnamefont {A.}~\bibnamefont {Glensk}}, \bibinfo
  {author} {\bibfnamefont {F.}~\bibnamefont {K\"ormann}}, \bibinfo {author}
  {\bibfnamefont {J.}~\bibnamefont {Neugebauer}}, \ and\ \bibinfo {author}
  {\bibfnamefont {R.~C.}\ \bibnamefont {Reed}},\ }\href@noop {} {\bibfield
  {journal} {\bibinfo  {journal} {Phys. Rev. B}\ }\textbf {\bibinfo {volume}
  {97}},\ \bibinfo {pages} {214106} (\bibinfo {year} {2018})}\BibitemShut
  {NoStop}%
\bibitem [{\citenamefont {Grabowski}\ \emph {et~al.}(2009)\citenamefont
  {Grabowski}, \citenamefont {Ismer}, \citenamefont {Hickel},\ and\
  \citenamefont {Neugebauer}}]{Gra2009}%
  \BibitemOpen
  \bibfield  {author} {\bibinfo {author} {\bibfnamefont {B.}~\bibnamefont
  {Grabowski}}, \bibinfo {author} {\bibfnamefont {L.}~\bibnamefont {Ismer}},
  \bibinfo {author} {\bibfnamefont {T.}~\bibnamefont {Hickel}}, \ and\ \bibinfo
  {author} {\bibfnamefont {J.}~\bibnamefont {Neugebauer}},\ }\href {\doibase
  10.1103/PhysRevB.79.134106} {\bibfield  {journal} {\bibinfo  {journal} {Phys.
  Rev. B}\ }\textbf {\bibinfo {volume} {79}},\ \bibinfo {pages} {134106}
  (\bibinfo {year} {2009})}\BibitemShut {NoStop}%
\bibitem [{\citenamefont {Grabowski}\ \emph {et~al.}(2015)\citenamefont
  {Grabowski}, \citenamefont {Wippermann}, \citenamefont {Glensk},
  \citenamefont {Hickel},\ and\ \citenamefont {Neugebauer}}]{GWG2015}%
  \BibitemOpen
  \bibfield  {author} {\bibinfo {author} {\bibfnamefont {B.}~\bibnamefont
  {Grabowski}}, \bibinfo {author} {\bibfnamefont {S.}~\bibnamefont
  {Wippermann}}, \bibinfo {author} {\bibfnamefont {A.}~\bibnamefont {Glensk}},
  \bibinfo {author} {\bibfnamefont {T.}~\bibnamefont {Hickel}}, \ and\ \bibinfo
  {author} {\bibfnamefont {J.}~\bibnamefont {Neugebauer}},\ }\href@noop {}
  {\bibfield  {journal} {\bibinfo  {journal} {Phys. Rev. B}\ }\textbf {\bibinfo
  {volume} {91}},\ \bibinfo {pages} {201103} (\bibinfo {year}
  {2015})}\BibitemShut {NoStop}%
\bibitem [{\citenamefont {Fultz}(2010)}]{Ful2010}%
  \BibitemOpen
  \bibfield  {author} {\bibinfo {author} {\bibfnamefont {B.}~\bibnamefont
  {Fultz}},\ }\href@noop {} {\bibfield  {journal} {\bibinfo  {journal} {Prog.
  Mater. Sci.}\ }\textbf {\bibinfo {volume} {55}},\ \bibinfo {pages} {247}
  (\bibinfo {year} {2010})}\BibitemShut {NoStop}%
\bibitem [{\citenamefont {Freysoldt}(2017)}]{Freysoldt2017}%
  \BibitemOpen
  \bibfield  {author} {\bibinfo {author} {\bibfnamefont {C.}~\bibnamefont
  {Freysoldt}},\ }\href {\doibase 10.1016/j.commatsci.2017.03.001} {\bibfield
  {journal} {\bibinfo  {journal} {Comp. Mater. Sci.}\ }\textbf {\bibinfo
  {volume} {133}},\ \bibinfo {pages} {71} (\bibinfo {year} {2017})}\BibitemShut
  {NoStop}%
\bibitem [{\citenamefont {Duff}\ \emph
  {et~al.}(2015{\natexlab{b}})\citenamefont {Duff}, \citenamefont {Finnis},
  \citenamefont {Maugis}, \citenamefont {Thijsse},\ and\ \citenamefont
  {Sluiter}}]{DUFF2015439}%
  \BibitemOpen
  \bibfield  {author} {\bibinfo {author} {\bibfnamefont {A.~I.}\ \bibnamefont
  {Duff}}, \bibinfo {author} {\bibfnamefont {M.}~\bibnamefont {Finnis}},
  \bibinfo {author} {\bibfnamefont {P.}~\bibnamefont {Maugis}}, \bibinfo
  {author} {\bibfnamefont {B.~J.}\ \bibnamefont {Thijsse}}, \ and\ \bibinfo
  {author} {\bibfnamefont {M.~H.}\ \bibnamefont {Sluiter}},\ }\href {\doibase
  https://doi.org/10.1016/j.cpc.2015.05.016} {\bibfield  {journal} {\bibinfo
  {journal} {Comput. Phys. Commun.}\ }\textbf {\bibinfo {volume}
  {196}},\ \bibinfo {pages} {439 } (\bibinfo {year}
  {2015}{\natexlab{b}})}\BibitemShut {NoStop}%
\bibitem [{mea()}]{meamfit}%
  \BibitemOpen
  \href@noop {} {}\bibinfo {note} {MEAMfit2 is an interatomic potential
  optimization package which can be obtained from STFC's Daresbury laboratory
  via the website https://
  www.scd.stfc.ac.uk/Pages/MEAMfit-v2.aspx}\BibitemShut {NoStop}%
\bibitem{prelim}
P. Srinivasan, Y. Ikeda, B. Grabowski, J. Janssen, A. Shapeev, J. Neugebauer, and F. K{\"o}rmann (in preparation).
\bibitem{SFE}
X. Zhang, B. Grabowski, F. K{\"o}rmann, A. V. Ruban, Y. Gong, R. C. Reed, T. Hickel, and J. Neugebauer, Phys. Rev. B {\bf 98} , 224106 (2018).
\bibitem{foot}
An attempt was made to optimize a reference-free modified EAM (RF-MEAM) potential, however due to the size of the potential parameter space we were unable to obtain an RF-MEAM potential which improved on the EAM potential. It is worth noting that this is an area of active research, with recent improvements by one of us (A.I.D.) to the underlying MEAMfit algorithm as well as, e.g., consideration of pre-converged binary and ternary potentials as starting points, likely to render such optimizations feasible in the near future.
\bibitem [{\citenamefont {Janssen}\ \emph {et~al.}(2018)\citenamefont
  {Janssen}, \citenamefont {Surendralal}, \citenamefont {Lysogorskiy},
  \citenamefont {Todorova}, \citenamefont {Hickel}, \citenamefont {Drautz},\
  and\ \citenamefont {Neugebauer}}]{pyiron-paper}%
  \BibitemOpen
  \bibfield  {author} {\bibinfo {author} {\bibfnamefont {J.}~\bibnamefont
  {Janssen}}, \bibinfo {author} {\bibfnamefont {S.}~\bibnamefont
  {Surendralal}}, \bibinfo {author} {\bibfnamefont {Y.}~\bibnamefont
  {Lysogorskiy}}, \bibinfo {author} {\bibfnamefont {M.}~\bibnamefont
  {Todorova}}, \bibinfo {author} {\bibfnamefont {T.}~\bibnamefont {Hickel}},
  \bibinfo {author} {\bibfnamefont {R.}~\bibnamefont {Drautz}}, \ and\ \bibinfo
  {author} {\bibfnamefont {J.}~\bibnamefont {Neugebauer}},\ }\href@noop {}
  {\bibfield  {journal} {\bibinfo  {journal} {Comp. Mater. Sci.}\ } (\bibinfo
  {year} {2018})}\BibitemShut {NoStop}%
\bibitem{py}
http://pyiron.org.
\end{thebibliography}

\end{document}